\documentclass[useAMS,usenatbib]{mn2e}

\usepackage{graphicx}

\title[Stellar Populations in disks, bulges and ellipticals]{The growth of disks and bulges during hierarchical galaxy formation - II: metallicity, stellar populations and dynamical evolution.}

\author[C. Tonini et al.]
{C. Tonini$^{1}$
\thanks{E-mail:chiaratonini19@gmail.com},
S. J. Mutch$^{1}$,
J. S. B. Wyithe$^{1}$, 
D. J. Croton$^{2}$
\\
$^{1}$School of Physics, University of Melbourne, Parkville, 3010 VIC, Australia\\
$^{2}$Centre for Astrophysics and Supercomputing, Swinburne University
of Technology, Hawthorn, VIC 3122, Australia\\
}

\begin{document}

  

\maketitle

\begin{abstract}

We investigate the properties of the stellar populations of model galaxies as a function of galaxy evolutionary history and angular momentum content. We use the new semi-analytic model presented in Tonini et al. (2016). This new model follows the angular momentum evolution of gas and stars, providing the base for a new star formation recipe, and treatment of the effects of mergers that depends on the central galaxy dynamical structure. We find that the new recipes have the effect of boosting the efficiency of the baryonic cycle in producing and recycling metals, as well as preventing minor mergers from diluting the metallicity of bulges and ellipticals. The model reproduces the stellar mass - stellar metallicity relation for galaxies above $10^{10}$ solar masses, including Brightest Cluster Galaxies. Model disks, galaxies dominated by instability-driven components, and merger-driven objects each stem from different evolutionary channels. These model galaxies therefore occupy different loci in the galaxy mass-size relation, which we find to be in accord with the Atlas 3D classification of disk galaxies, fast rotators and slow rotators. We find that the stellar populations' properties depend on the galaxy evolutionary type, with more evolved stellar populations being part of systems that have lost or dissipated more angular momentum during their assembly history.  
\end{abstract}


\section{Introduction}

The connection between dynamical signatures of the galaxy hierarchical assembly history such as dynamical structure and stellar kinematics, and the galaxy stellar population properties is at the core of our physical understanding of galaxy formation. 
Data of so-called ''early-type'' galaxies from surveys like ATLAS-3D (Cappellari et al. 2011, Emsellem et al. 2011) and other careful, in-depth observations (see a review by Kormendy \& Kennicutt 2004) show this to be a class of galaxies that is homogeneous in terms of photometric properties but with a variety of distinct dynamical structures and kinematic signatures. These include for instance angular momentum content, shape, and S\'{e}rsic index. Dynamics-based classifications, such as fast and slow rotators (Emsellem et al. 2011) and classic and secular/pseudo-bulges (Kormendy \& Kennicutt 2004), allow us to unveil the physical processes that leave an imprint on galaxy properties, and which can be investigated with theoretical models in the hierarchical clustering scenario. 

In Tonini et al. (2016; hereafter Paper I) we presented a new semi-analytic model which is an evolution of the model SAGE (Semi-Analytic Galaxy Evolution, Croton et al. 2016), in which we introduced a recipe to follow the angular momentum evolution of gas and stars based on the mass accretion and assembly history, a new star formation recipe based on the gas disk angular momentum evolution, and a new recipe to treat disk instabilities, minor and major mergers. All these prescriptions were motivated by dynamical considerations and were used to investigate the balance between secular evolution and violent processes in the galaxy population, and how they impact galaxy observables, such as the mass-size relation of galaxies. 

In this work we investigate how secular evolution and violent processes affect the stellar populations in the model galaxies, and whether dynamical signatures and stellar population properties correlate with each other. In particular we study galaxies with varying angular momentum content, which in our model characterises different evolutionary histories, and analyse their metallicity, age, colours and mass-to-light ratio. These properties probe different timescales of galaxy evolution, from fossil records of the entire history to the most recent short-term activity.

In particular, the metal content of galaxies is a fundamental property originating from the baryonic cycle of gas cooling, star formation and stellar feedback. Galaxy metallicity therefore represents a powerful tool to assess our modeling of galaxy evolution. 
While semi-analytic models often use the observed stellar mass - gas metallicity relation to either tune free parameters or check the model performance (Croton et al. 2006, 2016, Guo et al. 2011, 2013, Lu et al. 2011, 2015, De Lucia et al. 2004, De Lucia \& Blaizot 2007, Yates et al. 2013, Somerville et al. 2008, 2012)
the stellar metallicity is used less often (Lu et al. 2014, Ma et al. 2016). This is mainly because measurements of galaxy stellar metallicities are more affected by systematic uncertainties than gas metallicities (Gallazzi et al. 2005). In general semi-analytic models and hydrodynamical simulations have less success in reproducing stellar metallicity than gas metallicity (as discussed in Lu et al. 2014, Ma et al. 2016). 

However, the stellar metal content provides a very important insight into the evolution of galaxies. On the one hand, stellar metallicity is a property that can be produced across all galaxy types, irrespective of gas content, and therefore its variations within the galaxy population can be investigated for both star forming and quiescent galaxies, as well as in relation to morphological and dynamical properties, and in all environments.
On the other hand, while the gas metallicity represents an instantaneous snapshot of the new chemical state of a galaxy, the stellar metallicity is a record of the stellar population build-up across the whole star formation history (Ma et al. 2016). 

Stellar metallicity is a manifestation of the efficiency of the baryonic cycle, i.e. the interplay between star formation, gas inflows and outflows, and gas recycling, but also of the intensity and duration in time of the star formation, i.e. the shape of the star formation history. Therefore variations of metallicity across galaxies evolved through violent and secular processes will provide a physical link between assembly history and star formation history. In this work we explore this connection, and characterise how stellar metallicity, age, mass-to-light ratio and colours can be used as tracers of galaxy assembly.

To better understand the impact of our dynamical recipes on the stellar population observables, we do not tune any model parameters to match data on metallicity or other stellar population indicators (in Paper I we tuned the model to the $z=0$ stellar mass function). We compare our model's predictions with observational data as well as with the model SAGE (Croton et al. 2016) which was run on the same merger trees, to better isolate the effect of our new recipes.

The paper is organised as follows: in Section 2 we briefly present the relevant model recipes (but we defer the reader to Paper I for a more detailed description). In Section 3 we present the model predictions on the metallicity distribution in the galaxy population, in connection with the galaxy star formation history. In Section 4 we present a comparison of the model galaxy evolutionary types with data from ATLAS 3D, and discuss the stellar population properties (age, colours, mass-to-light ratios as well as metallicity) as depending on galaxy angular momentum content and evolutionary history. In Section 5 we discuss our results and in Section 6 we conclude. Throughout the paper, we adopt a value of the Hubble parameter of $h = 0.7$, and photometric magnitudes are in the AB system.

\section{Model overview}

The semi-analytic model we use is described in detail in Tonini et al. (2016a, Paper I), and here we present a brief summary of the relevant recipes. 
It is an evolution of the model SAGE (Croton et al. 2016, Lu et al. 2014) run on the Millennium dark matter simulation (Springel et al. 2005), where we adopt a new prescription for calculating the angular momentum evolution of the galaxy mass components. On this prescription we also base a new star formation law, and a new recipe for dealing with minor mergers and disk instabilities. On the other hand, the halo merger trees on which the model is run, gas cooling, gas stripping, feedback recipes, stellar Initial Mass Function and metal yields remain unchanged, facilitating a comparison with SAGE and allowing us to study the impact of the new recipes. 

We treat each event in the galaxy history as a perturbation, including gas accretion, star formation and feedback, mergers and disk instabilities. Each event induces an evolution $\delta \vec{J}$ in the angular momentum of the gaseous and stellar components. Depending on the event, angular momentum can be also transferred between components (for instance with a disk instability). We refer the reader to Paper I for details, but we briefly outline here the gas structural evolution, which determines the star formation. 
 
The cold gas mass that constitutes the galaxy gaseous disk grows from the cooling of hot gas trapped into the dark matter potential well (heated at the halo virial temperature) and from the accumulation of gas from infalling satellites. The cold gas gets depleted during star formation itself. Each of these events induces a perturbation in both mass and angular momentum of the cold gas:
\begin{eqnarray}
\delta M_{\rm{gas}} & = & \dot{M}_{\rm{cool}} \delta t - \dot{M}_{*} \delta t + M_{\rm{sat, gas}}~,  \\
\delta \vec{J}_{\rm{gas}} & = & \delta \vec{J}_{\rm{gas, cooling}} + \delta \vec{J}_{\rm{gas,sat}} + \delta \vec{J}_{\rm{gas,SF}} ~.
\label{mass_and_j}
\end{eqnarray}
For gas cooling from the halo or coming from satellites, the angular momentum variation is modeled as $\delta \vec{J}_{\rm{gas,i}} = \dot{M}_i (\vec{J_{\rm{DM}}} / M_{\rm{DM}}) \delta t$, implying that the incoming gas is in dynamical equilibrium with the halo just before the accretion. In the case of star formation, $\delta \vec{J}_{\rm{gas,SF}} = - \delta \vec{J}_{*}$. 

Based on gas mass and angular momentum, we calculate the gas density profile, assuming it to be exponential (see also Guo et al. 2011): $\Sigma(r) = \Sigma_{\rm{0}} \ \rm{exp}(-r/R_{\rm{D,gas}})$. The central density is defined as $\Sigma_{\rm{0}} = M_{\rm{gas}}/(2 \pi R_{\rm{D,gas}}^2 )$, and the disk scale-length is given by:
\begin{equation}
 R_{\rm{D,gas}} = \frac{J_{\rm{gas}} / M_{\rm{gas}}}{2 V_{\rm{max}}}~,
 \end{equation} 
where $V_{\rm{max}}$ is the peak rotation velocity of the dark matter halo. 
The gas radius evolves following perturbations in the angular momentum, but also when the cold gas disk loses mass at fixed angular momentum, as is the case following stellar feedback for instance. Then the remaining gas in the disk needs to redistribute its angular momentum, leading to a change in the disk radius. 

The gaseous disk structure in the model is self-consistently linked with the gas mass accretion history and dynamical state, and at any time the model calculates the gas density profile. We take advantage of this to produce a star formation law that depends on the disk structure, and therefore on the evolution of the angular momentum (see Paper I). 
We assume that only gas above the density threshold for star formation ($\Sigma_{\rm{crit}} = 10 M_{\odot}/pc^2$, Kormendy \& Kennicutt 2004)
can be turned into stars. From the density profile, we calculate the critical radius at which the density drops below the threshold, as:
\begin{equation}
R_{\rm{crit}} = R_{\rm{D,gas}} \cdot \rm{Log} \left( \frac{\Sigma_{\rm{0}}}{\Sigma_{\rm{crit}}} \right)~.
\label{Rcrit}
\end{equation}
The gas inside this radius is available for star formation. Its total mass is
\begin{equation}
M_{\rm{crit}} = M_{\rm{gas}}  \left[ 1 - \exp{ \frac{R_{\rm{crit}}}{R_{\rm{D,gas}}}} \cdot \left( 1 + \frac{R_{\rm{crit}}}{R_{\rm{D,gas}}}  \right)  \right]~.
\label{Mcrit}
\end{equation} 
A fraction of this gas, governed by an efficiency parameter $\epsilon = 0.25$, is turned into stars over a dynamical time defined as $t_{\rm{dyn}} = R_{\rm{crit}} / V_{\rm{vir}}$ (where $V_{\rm{vir}}$ is the halo virial velocity), so that the star formation rate is defined as:
\begin{equation}
SFR = \epsilon \cdot M_{\rm{crit}} / t_{\rm{dyn}}~.
\label{SFR}
\end{equation} 
Each new stellar population forms in dynamical equilibrium with the gas disk, and acquires its instantaneous specific angular momentum. The newly formed stars add onto the total stellar disk angular momentum with 
\begin{equation}
\delta \vec{J}_{*} = \delta M_{\rm{stars}} \cdot 2 V_{\rm{max}} R_{\rm{crit}} \frac{ \vec{J}_{\rm{gas}} }{ |\vec{J}_{\rm{gas}} |}~.
\label{jstars}
\end{equation} 
As with the gas, the stellar disk can evolve its angular momentum through stellar mass accretion from satellites, and disk instabilities that transfer mass into the bulge and redistribute the angular momentum in the disk. The disk structure is defined with an exponential profile with scale-length $R_{\rm{D}} = J_{*} / (M_{\rm{D}} \ 2 V_{\rm{max}} )$ (see Paper I). 

This formulation allows us to self-consistently determine the structure of gaseous and stellar disks and the star formation rate, thus anchoring the baryonic cycle to the galaxy dynamical evolution and assembly history. Notice that the angular momentum of gas and stars depends on the growth history of these components, while it is only marginally coupled with the instantaneous spin of the dark matter halo, contrary to SAGE and most models in the literature (see Paper I for a discussion on this point). 

Model bulges originate from mergers and disk instabilities. These are external perturbations caused by the merger tree (in the model disk instabilities are triggered by minor mergers), which represents environment at first order. 
The model calculates the angular momentum evolution and the dynamical structure of the galaxy post-merger, depending on the mass ratio, on the dynamics of the encounter, and crucially, on the current galaxy structure. This allows the galaxy to conserve its dynamical memory, with the consequence that the assembly history leaves observable imprints on the galaxy structure. 

This leads to the growth of two classes of bulges, depending on the assembly history. In what follows, we will refer to the merger trees as having two qualities, ''richness'' and ''speed''. The richness of the tree indicates the number of mass accretion events (other than quiescent gas cooling), while the speed indicates the rate of mass accretion over time (speed can vary in time as well). Both are indicators of galaxy environment.

Merger-driven bulges form in fast and rich merger trees. They originate in major mergers, and continue to grow through minor mergers at the high mass end. They are assumed to be pressure supported with no residual angular momentum. When there are no other stellar components, these objects are akin to elliptical galaxies. 

Instability-driven bulges form in rich and slow merger trees. They are the result of secular evolution, the sum of many minor perturbations over a long time.
These include minor mergers, fast cooling or tidal encounters, which trigger gravitational instabilities in the galaxy disk. When the disk becomes unstable, part of its stellar material dissipates angular momentum and sinks to the centre of the galaxy, growing the instability-driven bulge.
Note that minor mergers are absorbed by different components of the galaxy (disk, merger-driven or instability-driven bulge) depending on which one currently dominates the mass, and this affects the origin of the stellar populations of the bulge. So while in a disky galaxy the disk absorbs the satellite and becomes unstable, thus shedding disk stars into the bulge, in a bulgy galaxy with little or no disk, the satellite material impacts the bulge directly (see Paper I for details).  
This implies that the evolution of the galaxy structure and the distribution of its stellar populations is determined by the past assembly history as much as by the current encounter (see also Fontanot et al. 2015, Kannan et al. 2015). 

Disk-dominated galaxies form in poor and slow merger trees, where the mass accumulation due to in-situ star formation dominates over external mass accretion, and perturbations are minimal.

In the new model the merger tree determines the angular momentum evolution, which in turns determines the galaxy structure and the star formation rate. In this way star formation history and assembly history are self-consistently linked.
By contrast SAGE at every timestep in the simulation recalculates the galaxy structure, by instantaneously assigning one unique spin value to the galaxy, equal to the dark matter halo spin, and sets the disk scale radius as a fixed fraction of the virial radius (following Mo et al. 1998). This radius and the total cold gas mass determine the instantaneous star formation rate. SAGE assumes a disk of constant density, and the gas mass in excess of a star formation threshold is turned into stars (with an efficiency parameter). Disk instabilities and both minor and major mergers contribute to the growth of only one type of bulge, with no modeling of its structure or radius (tension between models and data stemming from this approach is investigated in Wilman et al. 2013). 

\subsection{The production of metals}

Both in the new model and in SAGE, each new stellar population inherits the metals of the gas from which it forms. Using the Chabrier stellar Initial Mass Function (Chabrier 2003) the model then calculates the mass of new metals (defined as elements heavier than Helium) produced by each new stellar generation, as $M_{\rm{mets}} = p_Z \cdot M_{\rm{SF}}$, where $M_{\rm{SF}}$ is the mass of newly-formed stars and $p_Z=0.025$ is the yield parameter for each solar mass formed (as in Croton et al. 2016). With the assumption that most metals are produced by massive stars, which are relatively short-lived compared to the average simulation timestep ($\sim 100 - 300$ Myr), the newly-produced metals are instantaneously returned to the cold gas phase, and are available for the next generation of stars. 

The metals produced by each stellar generation are put back into the cold gas component, and instantaneously mixed. If the metal-enriched gas remains in the cold disk, then it is available for the next stellar generation, and both the stellar and the gaseous disk self-enrich recursively. However, supernova feedback (and in the most massive galaxies, AGN feedback) reheats and/or removes part of the cold gas (see Tonini et al. 2016a and Lu et al. 2014 for details of feedback model parameters; see also Croton et al. 2016). The gas lost to the cold disk ends up in the hot component, carrying its metals with it. The metallicity of the hot component is determined by the balance between gas heated from the disk and pristine gas accreted from the large scale structure. Subsequent cooling of the hot component brings its metals back into the cold disk. If the feedback is strong enough, part of the gas is expelled from the galaxy, and remains in a reservoir bound to the galaxy halo, to be reincorporated at later times (see Croton et al. 2016). 

\begin{figure*}
\includegraphics[scale=0.6]{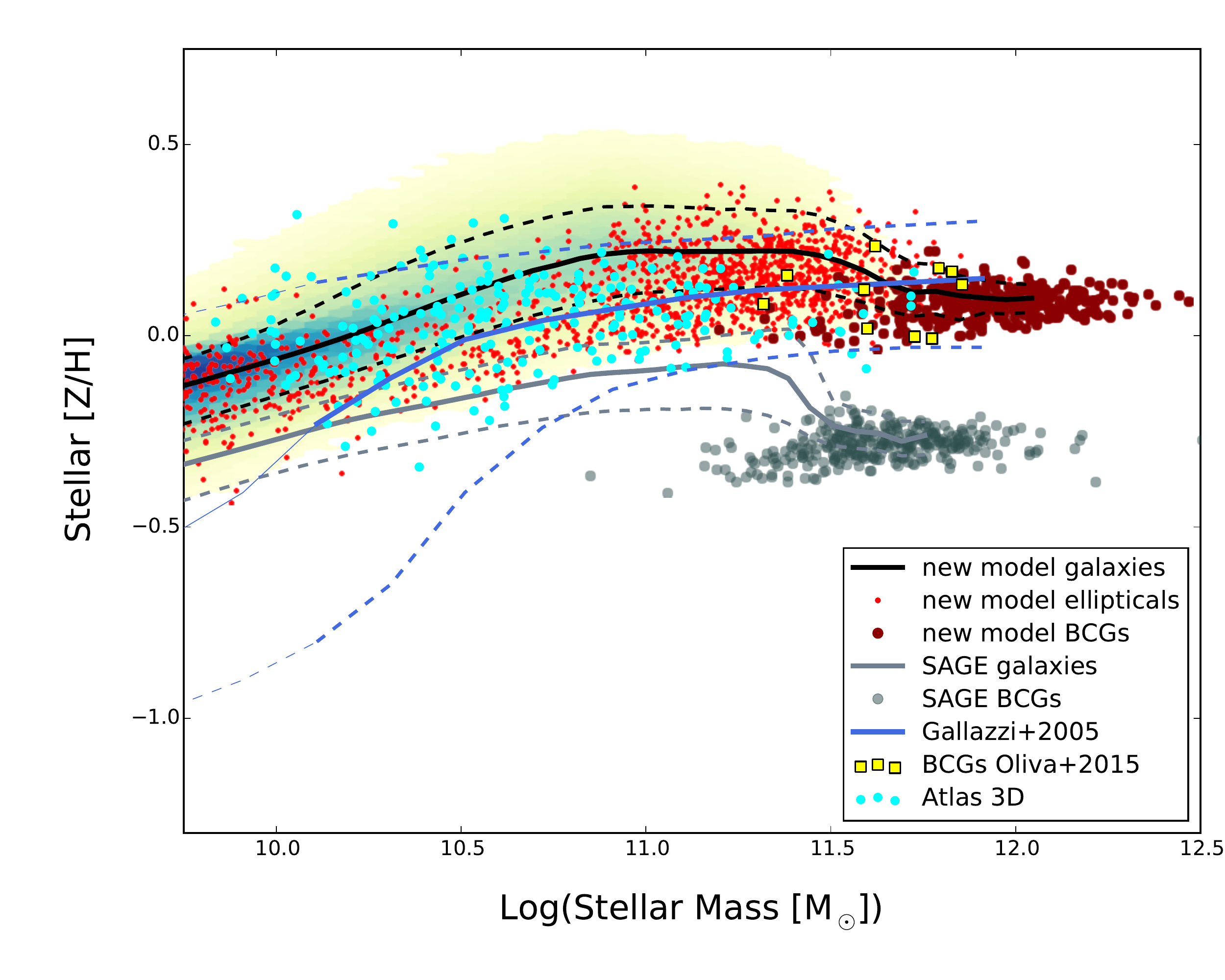}
\caption{The stellar mass - stellar metallicity relation. The model central galaxy population is represented with the blue-to-yellow 2D histogram and the black lines, solid for the median and dashed for the 68\% confidence levels. In red we show a random sample (for plotting clarity) of model elliptical galaxies (selected as merger-driven objects). In crimson circles we show the model Brightest Cluster Galaxies (BCGs). The observed Atlas 3D sample is represented with cyan circles, while the observed BCGs from Oliva-Altamirano et al. (2015) are represented with yellow squares. The blue solid line represents the median of the observed relation by Gallazzi et al. (2005), with the dashed lines representing the 68\% confidence levels. The grey lines represent the median and 68\% confidence levels of the model SAGE (Croton et al. 2016), and the grey circles represent SAGE BCGs.} 
\label{mass-metallicity}
\end{figure*}

In addition, the gaseous and stellar metallicity are affected by mergers. In general, a merger (major and minor) dilutes the central galaxy metallicity in both stars and gas. Because of the galaxy mass-metallicity relation, the incoming object being less massive also tends to be less metal-rich (Tonini, 2013). The merger remnant total stellar metallicity is a mass-weighted average of those of the incoming progenitors and of the new stellar component formed in the burst of star formation. The latter metallicity is determined by the gas present at the time of the merger. 

While merger-driven bulges acquire their metallicity as described above, the model instability-driven bulges are not merger remnants but are composed instead of disk material. Their stellar populations therefore show a continuity in physical properties with the disk itself, including colours, ages and metallicities. 

The model records the star formation history of each galaxy, including stellar mass and metals for each stellar generation, and in post-processing we interpolate it on the timegrid of the Conroy et al. (2009) synthetic stellar populations. We use the spectro-photometric model described in Tonini et al. 2012 (see also Tonini et al. 2009, 2010) to calculate the galaxy photometry from the star formation history. The results presented here do not include dust extinction. We compare the model photometry with samples selected to be dust-free (ATLAS 3D) or samples where dust effects have been accounted for in the morphological classification (Gavazzi et al. 2010). In addition, we focus on SDSS $g,r,i$ bands which are not heavily obscured, and across which the slope of the Calzetti extinction curve is relatively shallow (Calzetti, 1997), thus not heavily affecting the colours.

The results presented here are for a galaxy population with total stellar mass above $\rm{Log}(M_{\rm{star}}/M_{\odot}) > 9$, in a volume of $500 Mpc^3$ comoving. 
Results are presented for model central galaxies only. When model galaxies become satellites, they are prevented from further cooling of gas, and are just allowed to form stars from their remaining cold gas reservoirs, until exhaustion. While this recipe is more reasonable in clusters (which are however relatively rare objects), it is too drastic for galaxy groups, where it artificially reddens the galaxy colours. Group galaxies are at first order better represented by model centrals. 
The resulting model population comprises $\sim$4 million objects.

\section{Metallicity}

As discussed in Paper I, we can use the relative contribution of disk, merger-driven bulge and instability-driven bulge to the total stellar mass, and the fact that each of these components grow on different timescales, to reconstruct (statistically) the assembly history of the galaxy. Now we investigate whether the growth of these components produces other signatures, this time in the galaxy stellar populations, that can be used as additional diagnostic tools to reconstruct the galaxy history. 

\subsection{The stellar mass - stellar metallicity relation}

Fig.(\ref{mass-metallicity}) shows the stellar mass $vs$ stellar metallicity relation ($[Z/H]$ in solar units, where 0 is the solar value). The general model galaxy population is represented by the blue-to-yellow 2D histogram and the black lines, solid for the median and dashed for the 68\% confidence levels. With red circles we show the model elliptical galaxies, selected as merger-driven objects (for plotting clarity we only select a random sample of $\sim 10^5$ objects above $\rm{Log}(M_{\rm{star}}/M_{\odot}) > 9$). 
In crimson circles we highlight the model Brightest Cluster Galaxies (BCGs), selected as the central galaxies in the 300 most massive halos in the simulation (see also Tonini et al. 2012). The model is compared with the following data: BCGs from Oliva-Altamirano et al. (2015) with yellow squares, the Atlas 3D sample with cyan circles, and the observed stellar mass - stellar metallicity relation by Gallazzi et al. (2005). The latter is represented through its median (blue solid line) and the 68\% confidence levels (blue dashed lines). We note that their sample is affected by poor statistics at masses below $\rm{Log}(M_{\rm{star}}/M_{\odot}) = 10$ (see Fig. 6 in Gallazzi et al. 2005). 

The model agrees fairly well with the observed data. For masses $\rm{Log}(M_{\rm{star}}/M_{\odot}) > 10$ the model and the Gallazzi et al. (2005) observed relation agree within $1 \sigma$ of each other, and have the same slope. The model BCGs at the high mass end show a slight dip in metallicity, and show very good agreement with the observed BCGs of Oliva-Altamirano et al. (2015). The model also agrees with the data from Atlas 3D (which tend to be more metal-rich than the Gallazzi et al. 2005 sample), both with the general galaxy population and with the merger-driven objects only (red). Atlas 3D galaxies are characterised by a spread in angular momentum and ellipticity (i.e. a spread in bulge/total ratio) so that they should be compared with both. 
We also note that the model merger-driven bulges show a tighter mass-metallicity relation than the overall model population (as will be discussed later on in more detail). 
Although the scatter in the data becomes very large below $\rm{Log}(M_{\rm{star}}/M_{\odot}) = 10$, other datasets (Panter et al. 2008, Woo et al. 2008) show a similar trend, so there is a tendency of the model to be too metal rich at low masses. Both these effects are most likely regulated by the supernovae feedback efficiency, which affects the ratio between pristine and enriched gas available for star formation. 

The tail of the model galaxies extends towards very high metallicities at intermediate masses. As seen in Paper I, these galaxies are for the most part very active massive disks with instability-driven bulges. The excess of metals can be attributed to the calibration of the baryonic cycle (star formation efficiency $vs$ feedback efficiency), or the rate of reintegration of gas previously expelled by feedback and finding its way back onto the galaxy. 

In Fig.(\ref{mass-metallicity}) we also plot the predicted relation from the model SAGE, with the grey lines representing the median and 68\% confidence levels (solid and dashed respectively). We plot the SAGE BCGs as grey circles. We select the SAGE BCGs in the same way as in our new model, and since the two models are run on the same merger trees, then by definition they are the central galaxies in the same dark matter halos.  
The difference in performance between the two models is attributable to our new star formation recipe and the treatment of minor mergers. In the mass range $\rm{Log}(M_{\rm{star}}/M_{\odot}) > 10$ our model produces similar results to Somerville et al. (2008, 2012; see also Lu et al. 2014). Notice that the new model tends to produce more very massive galaxies ($\rm{Log}(M_{\rm{star}}/M_{\odot}) > 11.7$) than SAGE. However both models have been calibrated with the Baldry et al. (2008) observed stellar mass function, and the difference between the two model stellar mass functions remains within the observational uncertainties (see Paper I).

SAGE assumes that cold gas settles into a disk of constant density, with the radius determined instantaneously as a fixed fraction of the product of the dark matter halo virial radius and spin parameter. 
If the total gas mass is larger than a mass threshold for a given radius and halo mass, the excess gas is turned into stars, with an efficiency parameter. In the new model instead, the gas density profile is determined by the gas angular momentum history and accretion history, so that variations in the halo structure impact only marginally (through the current gas inflow) on the disk structure. Given the density profile, we calculate the critical radius inside which the gas is above the density threshold for star formation. This gas mass is turned into stars, with an efficiency parameter (see Paper I).

\begin{figure*}
\includegraphics[scale=0.4]{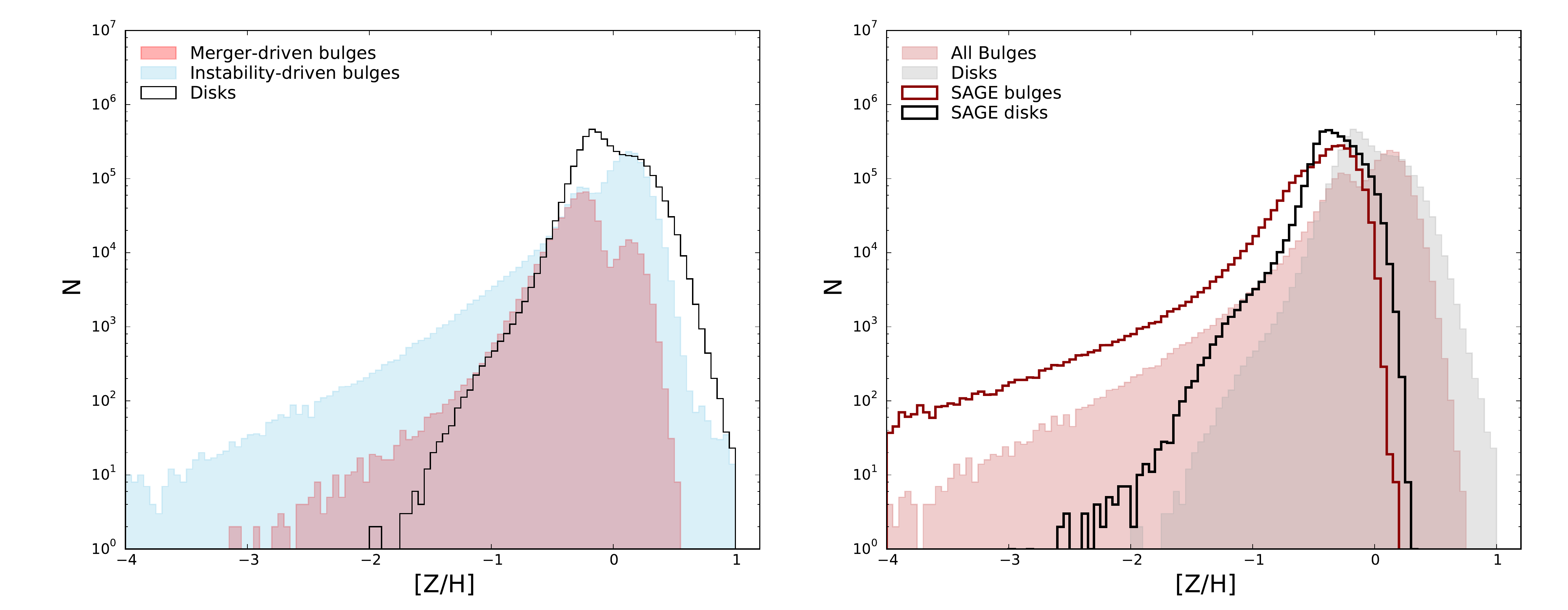}
\caption{Left panel: distribution of total metallicity [Z/H] in the merger-driven bulges, instability-driven bulges and disks, as indicated in the legend. Right panel: comparison between the metallicity distribution of the model disks and bulges (all bulges combined) with SAGE, as indicated in the legend.}
\label{mets}
\end{figure*}
 
The new model makes it harder for the gas disk to form stars. In other words the gas must accumulate for longer to reach the conditions for star formation. However when the density threshold is reached at a given radius, the gas inside that radius is more efficiently turned into stars. In addition, the ''pre-selection'' of the available gas through the density profile leads to a lesser burden on the supernovae feedback to keep the equilibrium in the baryonic cycle, so that the efficiency of SNe feedback can be lowered. 
As a result of the modified baryonic cycle and the near-dynamical decoupling from the halo, the gas system in the galaxy is more stable and self-enrichment (the re-incorporation of metals from one stellar generation to the next) is more effective. 
In addition to all this, in the new model the absorption of minor mergers in disk galaxies is regulated by the disk itself, and bulge growth is dominated by disk instabilities. Therefore minor mergers do not dilute the metallicity of bulges in disk galaxies. The effects of these two recipes trickle down the merger tree and affect all galaxies, including giant ellipticals and BCGs at the very high mass end of the population.

\subsection{The distribution of metals in disks and bulges}

The metal abundance in a galaxy is the result of its star formation history, which as we discussed in Paper I, is linked to its hierarchical assembly history. In this Section we investigate the metal content of the stellar mass components resulting from 3 different growth mechanisms: merger-driven bulges that are the product of major mergers (and mergers on spheroids), instability-driven bulges produced by minor mergers and disk instabilities, and disks formed through quiescent star formation. As the processes that develop these stellar mass components are different and directly relate to the environment where the galaxy evolves and the richness of the merger tree, we investigate if the mass component distributions in metallicity reflect these differences.

The left panel of Figure (\ref{mets}) shows the model total metallicity distribution $[Z/H]$ for the merger-driven (\textit{red}) and the instability-driven (\textit{blue}) bulges, as well as the disks (\textit{black line}). The right panel shows the comparison of the model metallicities with those predicted in SAGE. Since SAGE does not produce two distinct bulge populations, for the comparison the merger-driven and instability-driven bulges are grouped together. 

The most notable feature of this plot is the double peak in metallicity that occurs for all components. As will be investigated below, it is an effect driven by the transition of the galaxy population following the change of the dominant growth mechanism for bulges, as a function of galaxy mass. As shown in Paper I, at masses around $\rm{Log}(M_{\rm{star}}/M_{\odot}) > 10$ secular evolution starts to dominate bulge formation, and instability-driven bulges therefore inherit most of the metals formed in disks, until merger-driven bulges become dominant again at masses $\rm{Log}(M_{\rm{star}}/M_{\odot}) > 11.5$. This competition is at the origin of the double feature in the metallicity distribution.

We notice from the left panel that the metallicity distributions of disks and merger-driven bulges have peaks within $\sim$ 0.2 dex of each other, while instability-driven bulges peak at about $\sim 0.5$ dex higher metallicity. All components have a significant high-metallicity population (above solar), with disks and instability-driven bulges reaching $[Z/H] \sim 1$ and merger-driven bulges reaching $[Z/H] \sim 0.5$. 
All components show an asymmetric distribution, with a shorter high-metallicity tail and a longer low-metallicity one. Disks show the tightest distribution, not dropping below $[Z/H] \sim -2$, while the bulges, and especially those which are instability-driven, reach well beyond $[Z/H] \sim -3$.

The new model is much more efficient than SAGE in producing metals in disks and distributing them into bulges. The peak value for SAGE bulges and disks are respectively $\sim 0.5$ and $\sim 0.3$ dex lower than the new model, and they are more metal-poor than the new model across the whole metallicity range. The difference is most pronounced at the high-metallicity end, where only a small fraction of galaxies reach super-solar metallicities. In particular, contrary to the new model, SAGE does not produce metal-rich elliptical galaxies.  

Apart from occasional bursts of star formation accompanying violent events (such as gas-rich mergers), \textbf{the vast majority of metals ($>90\%$ on average)} are produced in disks through in-situ star formation. In Paper I we argue that the disk formation is a process that spans a few Gyrs. In this time interval, multiple stellar populations must be formed from gas cooling at a \textbf{relatively steady pace to maintain} the disk gravitational stability. The resulting disk population is self-enriched to a minimum $[Z/H] \sim -2$, with the vast majority of disks above $[Z/H] \sim -1$. At the other end of the range, disks are pushing the metal-rich boundary of the overall metallicity distribution towards higher and higher values. The metal-rich end shows no tail petering out, but rather the slope is steep and the cutoff sharp, indicating that a whole subpopulation of star-forming galaxies is driving this trend. 

Although instability-driven bulges form from disk material, their metallicity distribution is skewed compared to that of disks, with a peak at higher metallicities. Disk instabilities that lead to the formation of these bulges are more frequent at the most star-forming, high-mass end of the disk population (see Paper I), so the peak in metallicity of their distribution is driven by the high-metallicity end of the disk distribution. But while instability-driven bulges acquire their stellar populations from the disk, contrary to the disk they are not subject to direct gas cooling and accretion of satellites. Every minor merger statistically dilutes the metallicity of the disk that absorbs it, while the bulge is sheltered from this process. 
Depending on the merger tree, bulge and disk grow in metallicity at different speeds, with the disk being slowed down by incoming low-metallicity material, and secular processes selectively locking away metal rich populations in the bulge.

As will be discussed below, the extended low-metallicity tail is instead due to very low-mass systems, likely to have formed at high redshift. 

Disks, by virtue of their prolonged star formation histories that foster self-enrichment, dominate the high-metallicity tail of the galaxy population.  
Merger-driven bulges peak at a similar metallicity to that of disks, but at the high metallicity end they show a ''delay'' of about $0.4 - 0.5$ dex. This metallicity gap stems from two causes. 

The first is that the regions of peak activity of major mergers and in-situ star formation are mismatched. Or in other words, at any given time, merging galaxies and galaxies with the highest star formation rates belong to different merger trees.
In Paper I we show how galaxies with high star formation activity have massive disks and develop instability-driven bulges. These are the most metal-rich galaxies, and statistically, their merger trees do not contain a major merger after $z \sim 1$. 
In fact, notice that the metallicity distribution of merger-driven and instability-driven bulges do not have the same shape. This is an additional indication that galaxies dominated by instability-driven bulges are not the primary source of progenitors for major mergers. 

Since galaxies with instability-driven bulges will likely never be involved in a major merger, then the progenitors of merger-driven bulges must be found in the rest of the population. These are disk galaxies or ellipticals with star formation rates and metallicities closer to mean values (hence the peaks of the distributions are similar), while it is extremely rare that they belong to the very high end of the metallicity distribution.

The second cause for the ''metallicity delay'' can be interpreted as a time delay, stemming from the fact that the populations of disks and ellipticals grow at different speeds. Major mergers are statistically suppressed, so the build-up of the merger-driven population is slow.
Since only a small fraction of galaxies become progenitors in a major merger, and therefore deposit their metals in a merger-driven object, then the merger-driven population trails behind the others in numbers, slowed down by the extreme unlikeliness of a major merger. Note that, given the relative positions of the peaks, at any given time there is always a tail of high metallicity disks that push the metallicity distribution and never merge.

\begin{figure}
\includegraphics[scale=1.2]{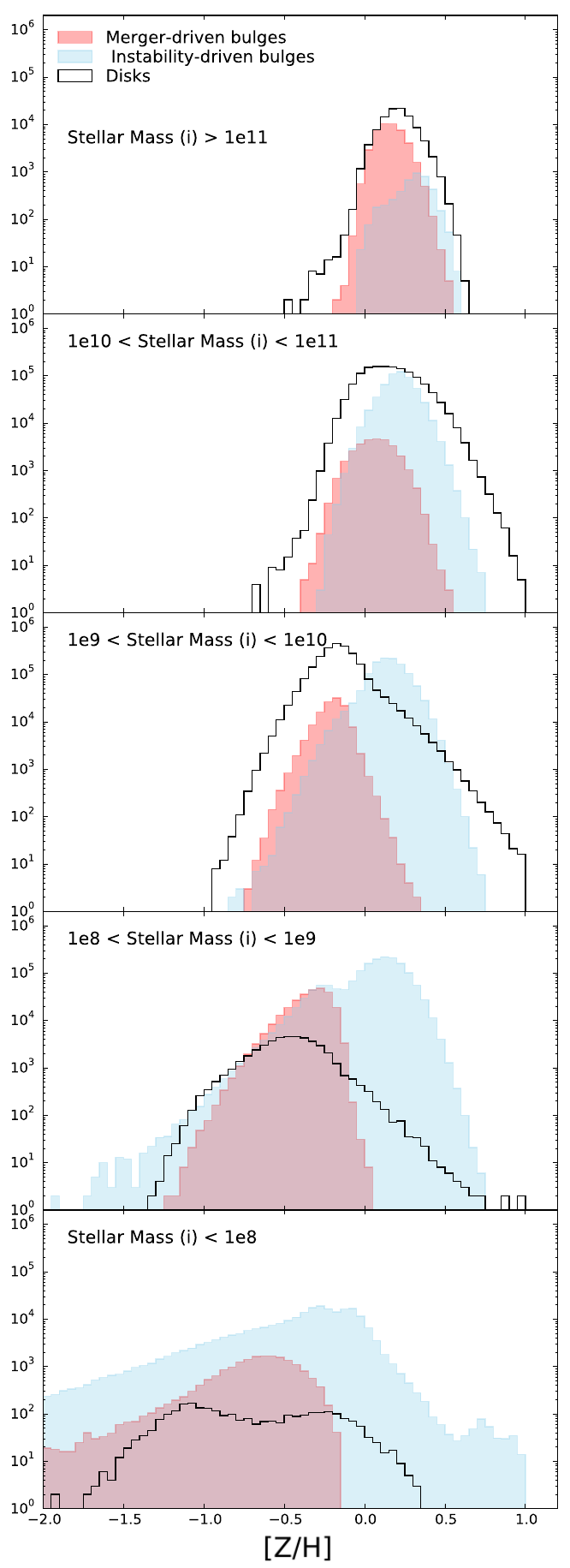}  
\caption{Distribution of total metallicity [Z/H] in the merger-driven bulges, instability-driven bulges and disks (as indicated in the legend), where each component is split in bins of stellar mass. The \textit{y-axis} represents the number of galaxies.}
\label{mets-multiple}
\end{figure}

\subsection{Scatter in the metallicity distribution}

Given that model galaxies follow a mass - metallicity relation, stellar mass is a major contributor to the width of the metallicity distribution. To investigate other sources of scatter, we split the model disks, merger-driven and instability-driven bulges in bins of mass. Figure (\ref{mets-multiple}) shows their metallicity distribution (colour-coded as in Fig.~(\ref{mets}) in 5 mass bins, referring to the stellar mass of the component itself, while the total galaxy stellar mass of the sample is again $\rm{Log}(M_{\rm{star}}/M_{\odot}) > 9$. Notice the \textit{x-axis} is zoomed in on $(-2 < [Z/H] < 1)$ for clarity. 

Fig.(\ref{mets-multiple}) shows that disks, merger-driven, and instability-driven bulges each follow a mass-metallicity relation, with the merger-driven bulge one (red) being the tighest and steepest. 
Notice that the double feature in the metallicity distribution in Fig. (\ref{mets}) was driven by the variation of galaxy mass. The trough in the distribution at around the solar metallicity value arises from the movement of the merger-driven and disk peaks 
at masses around $\rm{Log}(M_{\rm{star}}/M_{\odot}) \sim 10$, where a change in the shape of the distribution also appears. From Paper I, we know that the model galaxy population around this mass is undergoing a transition, in which secular evolution starts to dominate bulge growth, and disk galaxies switch from having merger-driven bulges to instability-driven bulges.

In merger-driven bulges, we find that mass and metallicity are correlated, and they are both anti-correlated with dynamical age (see Paper I). 
Following the growth of structures, the mass of major merger progenitors increases with passing time. Therefore, in the highest mass bin we find not only the most metal rich, but also the dynamically youngest elliptical galaxies. The lower mass bins contain dynamically old merger-driven objects, relics of a low-metallicity universe, that were formed in the early phases of galaxy assembly. Below masses of approximately $\rm{Log}(M_{\rm{star}}/M_{\odot}) \sim 10$, model merger-driven bulges can be found as both low-mass ellipticals and bulges inside disk galaxies. Above this limit, merger-driven bulges are almost exclusively elliptical galaxies, with no disks. Their abundance drops significantly relative to the other components, until galaxy masses reach $\rm{Log}(M_{\rm{star}}/M_{\odot}) \sim 11.5$ (see also De Lucia et al. 2012). The slope of their mass-metallicity relation flattens, because major mergers are removed from the regions of more intense star formation, and metals become locked in merger-driven bulges only later, when structure growth has collapsed these regions. 
This effect is stronger at the highest mass bin, which is the only mass regime where minor mergers are a non-negligible source of mass growth for elliptical galaxies (see Paper I). Because of the mass-metallicity relation, minor mergers tend to dilute the metallicity of the central galaxy by accreting low-mass satellites.
 
Merger-driven bulges only grow their stellar mass through assembly, which does not increase stellar metallicity (unless there is metal-rich gas involved in a merger). It can be argued that the regularity displayed by merger-driven bulges, and in particular their mass-metallicity relation, shows the workings of hierarchical clustering, which populates the denser regions of the universe with the most evolved stellar populations. 

Disks show a much larger scatter in metallicity at every mass bin. Somewhat counter-intuitively, in-situ star formation is a lot less precise than assembly via mergers in shaping the mass-metallicity relation. 
Environmental conditions (i.e. the richness of the merger tree) are very important in shaping the disk star formation history, which shows a large variety (in terms of duration and intensity). At total galaxy masses in the interval $10 \sim  \rm{Log}(M_{\rm{star}}/M_{\odot}) > 11.5$ disks find themselves in merger trees devoid of major mergers (see Paper I). As discussed in Paper I, this is where star formation is most easily sustained over long periods of time, facilitating self-enrichment and leading to high-metallicity tails in the distribution. 
Secular evolution dominates the galaxies in this mass range, leading to the formation of instability-driven bulges, which populate the most instensely active trees. 

At all masses the peak metallicity of instability-driven bulges is higher than that of disks, and their distribution is tighter except in the lowest mass bin. The trend between mass and metallicity is arguably marginal. 
Given a population of disks, the objects most likely to develop instability-driven bulges are the ones with the most intense tree activity, and the highest star formation rate. Therefore, these bulges tend to originate from the massive, metal-rich end of the disk population. This drives the peak of the instability-driven bulge distribution to higher metallicities than that of the disks. 

Contrary to the merger-driven case, in instability-driven bulges mass and dynamical age are not so clearly anti-correlated. Given the long timescales necessary for the growth of these objects (of the order of $6-8$ Gyr, see Paper I), their dynamical age is sensitive to the merger tree activity over time, i.e. the ''speed'' of the tree. For instance fast early growth or slow late growth can produce bulges of the same mass, but with very different dynamical ages, as well as very different stellar populations. Notice that the relative metallicity of disk and instability-driven bulge in the same galaxy reflects the differential speed of the merger tree in time: a fast and eventful early tree activity causes an early metal-poor bulge, followed by a steady increase in disk metallicity during the more quiescent galaxy evolutionary phases. On the other hand, later and steadier activity leads to a metal-rich bulge, that locks away the metals produced in the disk, safe from the dilution that the disk might experience.

The lowest mass bin shows a huge spread in metallicity. Each of these bulges and disks is small and not the dynamically dominant mass component in galaxies selected as $\rm{Log}(M_{\rm{star}}/M_{\odot}) > 9$. Disks at these masses are not very competitive in securing gas supply, nor in retaining it after supernovae feedback\footnote{All galaxies represented in the plots are centrals, so these effects do not include stripping of gas, which affect satellites.}. Merger-driven bulges at these masses are ancient and therefore very metal-poor. Given the delay in the formation of instability-driven bulges after a disk is established (see Paper I), at very low masses this type of bulge can either be very old and metal poor, or very young and belong to a massive metal-rich disk, hence the large metallicity spread. 

\bigskip

With a limited mass range in each bin, the spread in metallicity is caused by the variations in the assembly history. The accumulation of metals reflects the star formation history, the relative number of minor-to-major mergers and the relative incidence of mergers versus local star formation. These factors determine the richness and speed of the merger tree, and represent the environment of the galaxy. We now proceed to investigate how different evolutionary histories affect general stellar population properties and structural properties of the model galaxies. 

\section{Galaxy evolutionary types}

\begin{figure}
\includegraphics[scale=0.38]{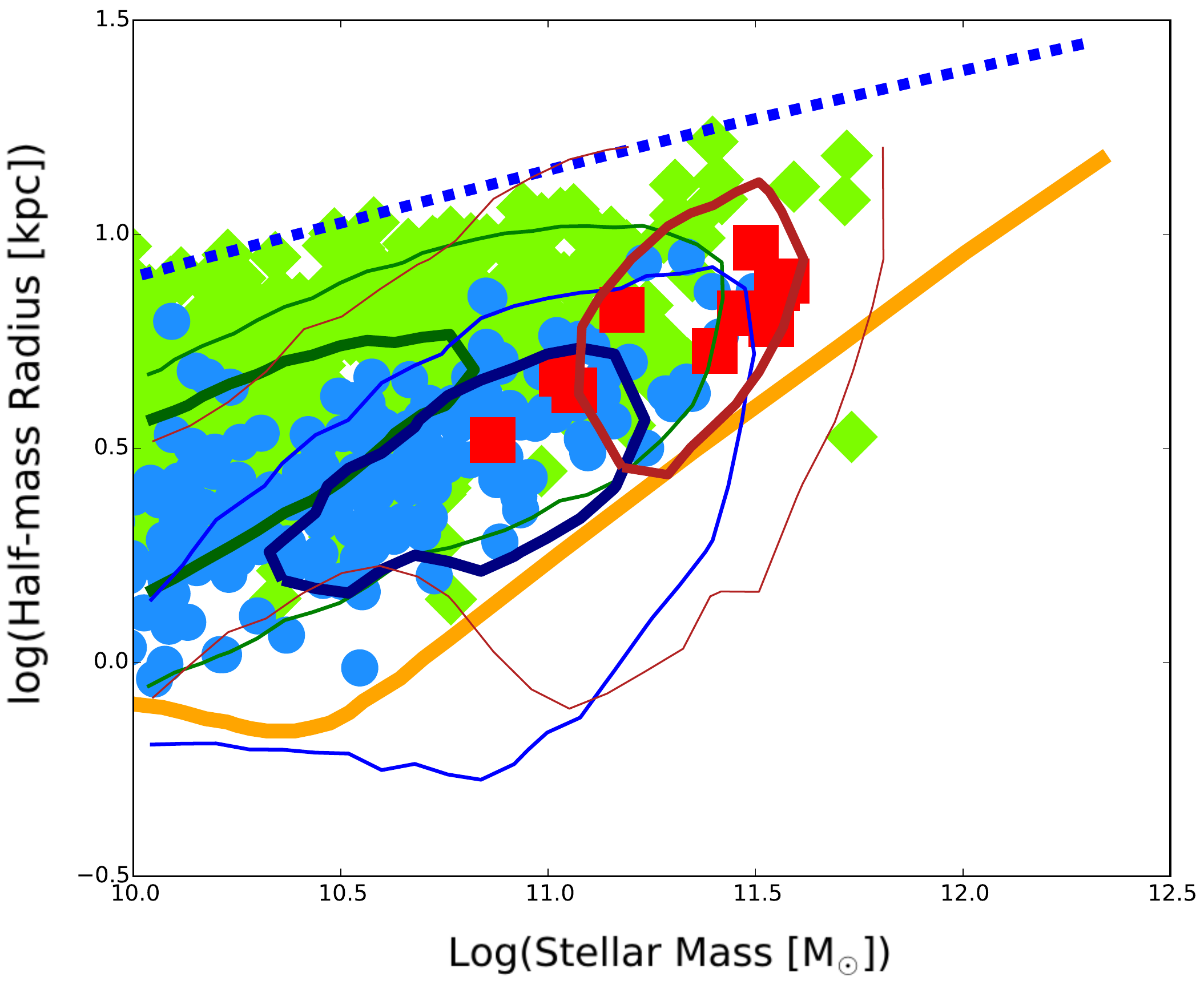}
\caption{Contours: the mass-size relation of the model galaxies, split by galaxy evolution mechanism. Red contours represent objects where the merger-driven component represents 60\% or more of the stellar mass, green contours represent objects where the disk represents more than 80\% of the stellar mass, and blue contours represent objects dominated by secular evolution, where the bulge is composed of 70\% or more by the instability-driven component and the disk is less than 80\% of the stellar mass. The contour lines represent the $1\sigma$ and $2\sigma$ confidence levels for each type. Data points: the ATLAS 3D galaxy sample. Green diamonds represent spirals, blue circles represent fast rotators, and red squares represent slow rotators, as defined in Cappellari (2013). The solid orange and dashed blue lines are taken from Figure 9 of Cappellari et al. (2013b).}
\label{atlas}
\end{figure}

Figure (\ref{atlas}) shows the model stellar mass - size relation, with size represented by the half-mass radius (see Paper I), where we have divided galaxies according to their ''evolutionary type''. Red contour lines represent objects where 60\% or more of the stellar mass is in the merger-driven bulge (including all elliptical galaxies). Green contour lines represent objects where 80\% or more of the stellar mass is in the disk, and which are therefore mainly assembled through quiescent star formation with minimal perturbations. Blue contour lines represent objects with bulges dominated by the instability-driven component (specifically, the bulge is more than 70\% instability-driven and the disk accounts for less than 80\% of the total mass), indicating a secular-like evolution dominated by perturbations and minor mergers. 
The thick and thin density contour lines represent the $1\sigma$ and $2\sigma$ confidence levels for each type. 
The data points represent the observed ATLAS 3D sample (Cappellari et al. 2011, Cappellari et al. 2013a), where green diamonds are classified as spiral galaxies, blue circles are classified as fast rotators, and red squares are classified as slow rotators (Emsellem et al. 2011). The orange solid line and the blue dashed line are taken from Cappellari et al. (2013, their figure 3), and represent the envelope of their sample. We find that the correspondence between the model galaxies behaviour and the observed Atlas 3D sample is very good. In particular, we find a correspondence in this plot between the Atlas 3D slow rotators and the model merger-driven objects, and between the Atlas 3D fast rotators and the model galaxies with growth dominated by secular evolution, which develop instability-driven bulges. Model disk galaxies also match the Atlas 3D spirals. 

Each event in the model galaxy evolution (star formation, accretion of gas, accretion of satellites, gas loss via feedback) produces an evolution of the angular momentum of all galaxy components (see Paper I). Therefore in the mass-size diagram, different galaxy evolutionary histories produces different sequences.  
For instance, model galaxies dominated by secular evolution (blue) are disky galaxies that have undergone a redistribution of their mass and angular momentum due to the formation of their instability-driven bulge. Disk instabilities lead to the migration and concentration of stars into the galaxy centre, with the consequent steepening of the total stellar density profile, and the decrease of the half-mass radius, compared with pure disks of the same mass. 

The size of model merger-driven objects is instead determined by the size of the progenitors and the orbital parameters of the merger encounter. The merger remnant is assumed to be dominated by dispersion, with no residual angular momentum after the encounter. These objects dominate the high-mass end of the galaxy population. 
At low masses, disks re-grow around ancient merger-driven objects, and the galaxy angular momentum evolves accordingly. 

The scatter in structural properties of model galaxies, including radius and bulge-to-disk ratio, depend on the variety of assembly histories. At a given mass these depend on the richness and speed of the merger tree, or in other words environment. 

\subsection{Stellar population properties}

\begin{figure}
\includegraphics[scale=0.3]{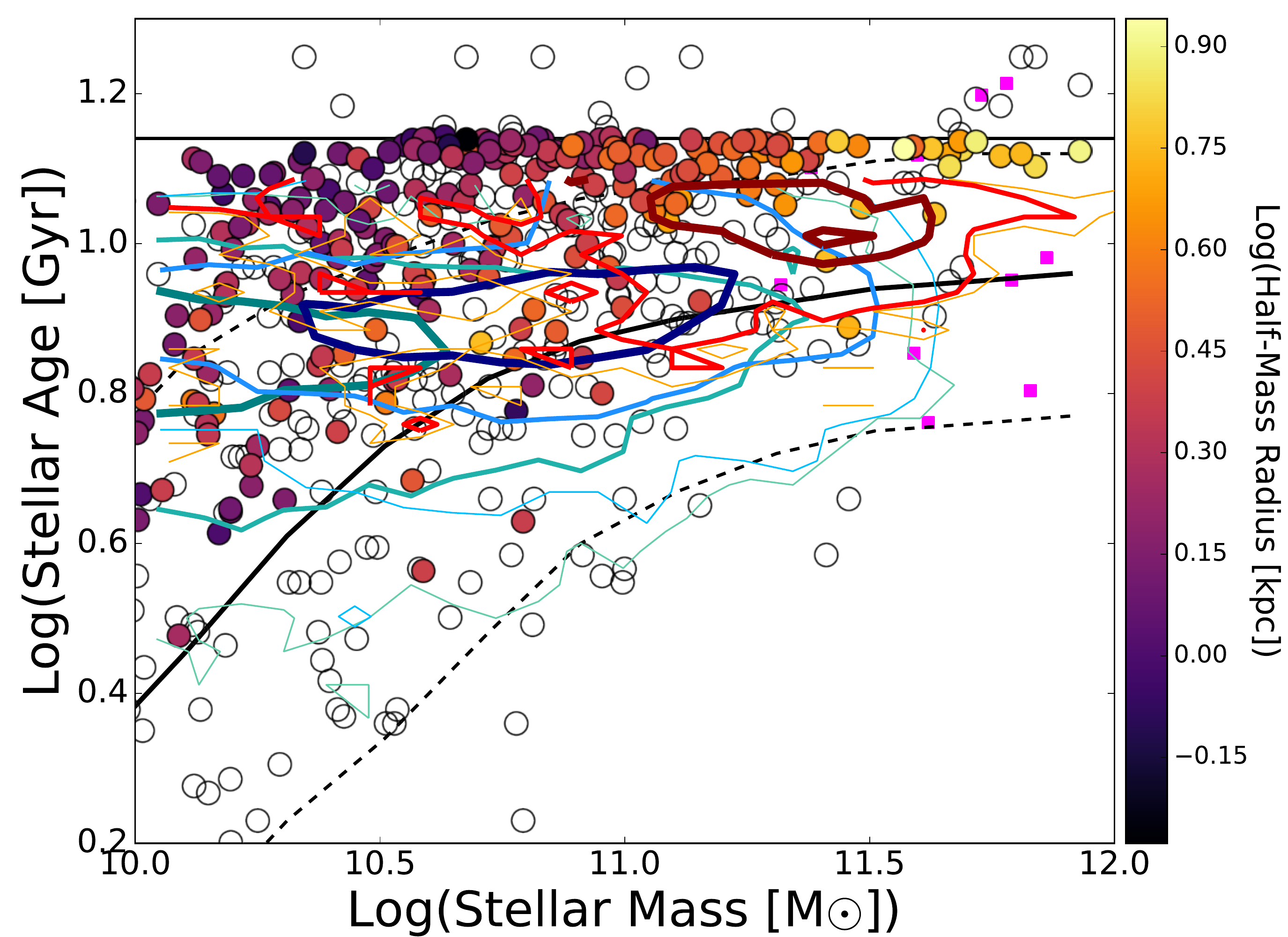}
\caption{Galaxy stellar ages as a function of stellar mass. Each evolutionary type of the model galaxies is represented by the coloured contours at the $1,2,3\sigma$ confidence levels, with contour colours as in Fig.~\ref{atlas}: shades of red for objects dominated by a merger-driven bulge, shades of blue for disky objects with bulges dominated by an instability-driven component, and shades of green for disk galaxies. The \textit{filled circles} represent Atlas 3D galaxies from McDermid et al. (2015), colour-coded in Log(effective radius) as in the \textit{color bar}. For these galaxies the ages are mass-averaged quantities over the recovered star formation history.
For comparison, the \textit{empty circles} represent ''SSP-equivalent'' ages recovered from Lick indices (for details see McDermid et al. 2015). The \textit{black solid and dashed lines} represent respectively the median and the 68\% confidence levels of the Gallazzi et al. (2005) dataset. The \textit{magenta squares} show the BCGs from Oliva-Altamirano et al. 2015. The horizontal line shows the age of the Universe.}
\label{ages}
\end{figure}

\begin{figure}
\includegraphics[scale=0.3]{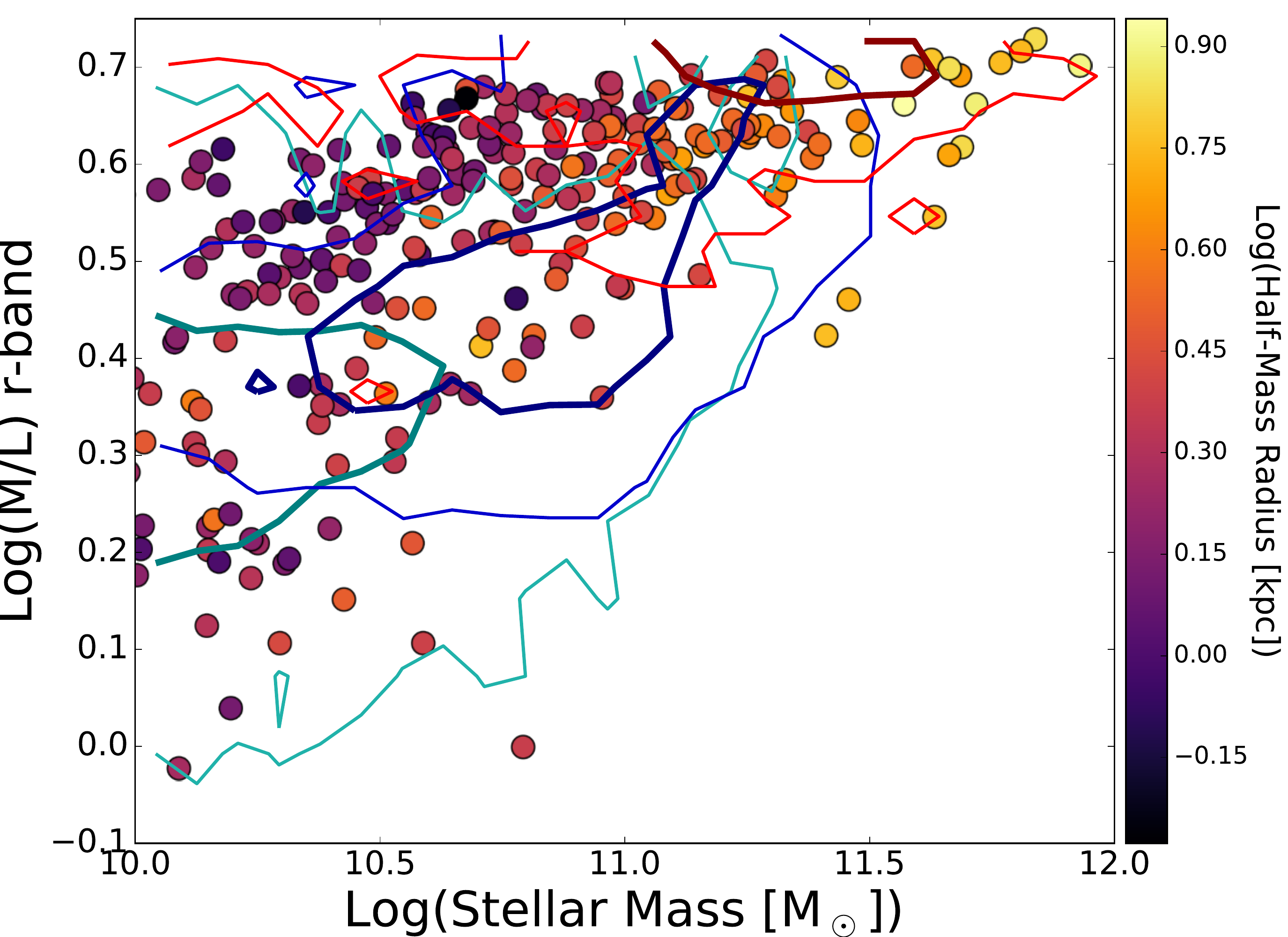}
\caption{Stellar mass-to-light ratio of the model galaxies, divided in evolutionary types with contour colours as in Fig.~\ref{atlas} and represented by the $1\sigma$ and $2\sigma$ contours. The \textit{filled circles} represent Atlas 3D galaxies from McDermid et al. (2015), colour-coded in Log(effective radius) as in the \textit{color bar}.}
\label{ML}
\end{figure}
  
\begin{figure}
\includegraphics[scale=0.3]{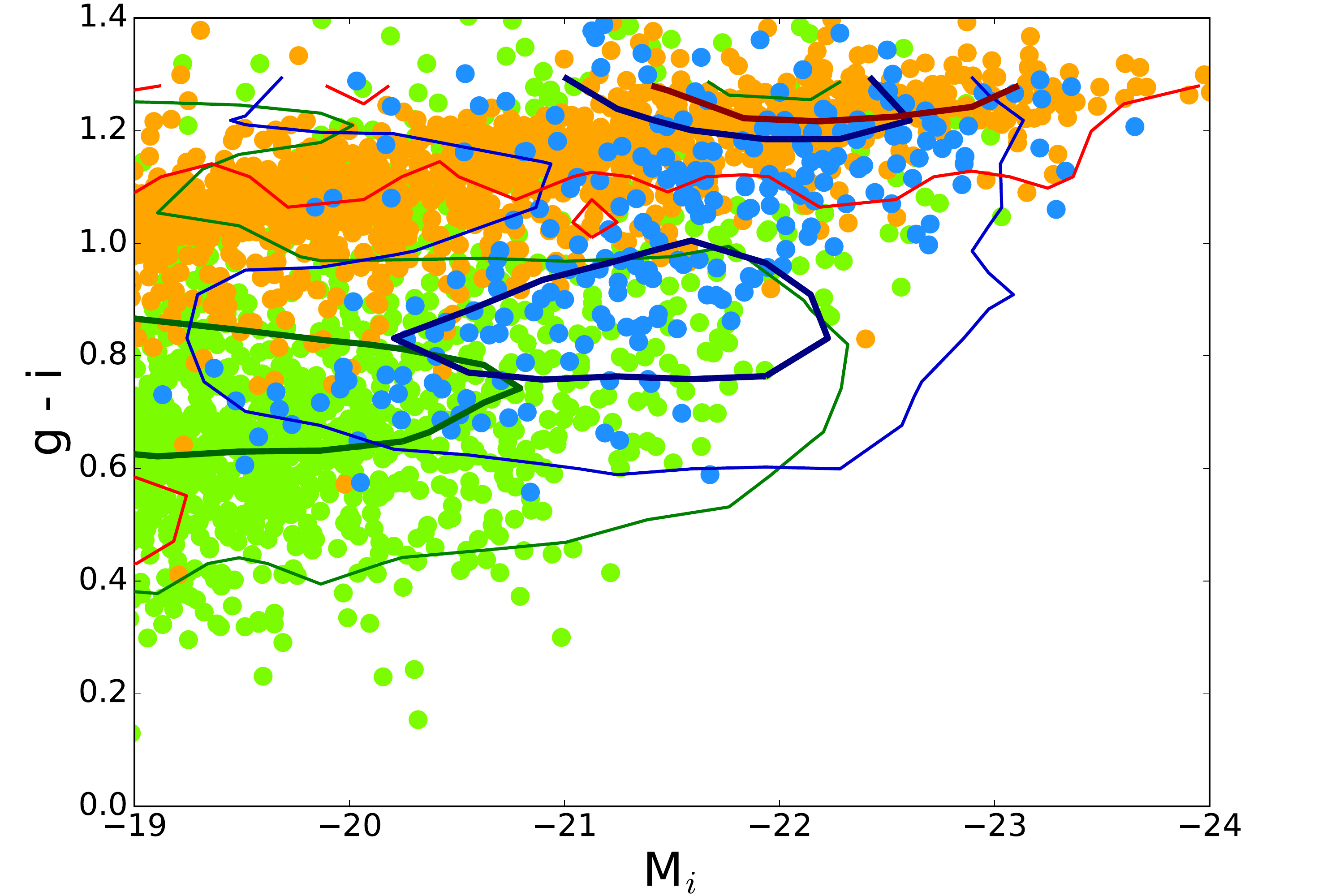}
\caption{The colour-magnitude diagram $(g-i)$ vs $M_i$ for the model galaxies, represented by the $1\sigma$ and $2\sigma$ contours for each evolutionary type, with contour colours as in Fig.~\ref{atlas}. The coloured circles represent the dataset of Gavazzi et al. (2010), divided according to visual morphology: \textit{red} for ellipticals and S0s, \textit{green} for disk galaxies (Sbc and later) and \textit{blue} for disk galaxies with prominent bulges (Sa-Sb).}
\label{colormag}
\end{figure}

The scatter in the model mass-size relation is due to different evolutionary paths, which result in differences in the galaxy structure. We can now investigate if such differences are mirrored by variations of the stellar populations in these galaxies. But first, we compare the model stellar population properties with the ATLAS 3D galaxies and with a larger SDSS sample by Gavazzi et al. (2010). 

Fig.~(\ref{ages}) shows the stellar ages as a function of the stellar masses. Model galaxies are divided into three evolutionary types as in Fig. (\ref{atlas}), and represented by $1,2,3 \sigma$ contours with the same colour-coding: shades of red for objects dominated by a merger-driven bulge, shades of blue for disky objects with bulges dominated by an instability-driven component, and shades of green for disk galaxies. 
The \textit{filled circles} represent the ATLAS 3D galaxies from McDermid et al. (2015), where ages are calculated as a mass-average over their recovered star formation histories. They are also colour-coded with the logarithm of their effective radius (see the colour bar). As the recovery of stellar age in observed galaxies is heavily model-dependent, 
McDermid et al. (2015) propose alternative popular methods, like ''SSP-equivalent'' ages recovered from Lick indexes, which here we show for comparison with the \textit{empty circles}. The \textit{magenta squares} show the BCGs from Oliva-Altamirano et al. (2015).
For reference, the \textit{black solid and dashed lines} represent respectively the median and the 68\% confidence levels of the Gallazzi et al. (2005) stellar mass - age relation for their SDSS sample (which includes all galaxy types), also obtained with a Lick indexes analysis. The horizontal line shows the age of the Universe.

The model does a good job of reproducing the ages of the ATLAS 3D galaxies, with galaxies dominated by merger-driven and instability-driven components. Model galaxies of decreasing angular momentum move from high mass and high age to low mass and low age, along the mass-age relation defined by the data. In the next Section we will analyse the trend of model ages with mass and radius, but for now we point out that ATLAS 3D galaxies show a tighter relation at the high-mass end, with massive and extended objects with very old stellar populations, and an increasing scatter in age towards lower masses, with a slight tendency of more compact objects to be older. 

The model lies well above the Gallazzi et al. (2005) relation at masses below $Log(M_{star}) < 10.5$. We argue that this discrepancy has two causes. On the one hand, the model ages are mass-averaged over all the stellar populations in the galaxy. Given the progression of the cosmic star formation rate with redshift, it is very rare to find galaxies in the simulation that form the bulk of their stars after $z=1$, while the Gallazzi et al. (2005) relation implies that the median age at $Log(M_{star}) = 10.5$ is $\sim$5 Gyr. While indeed we find that almost all galaxies have residual star formation down to very low redshifts, this normally accounts for only an incremental variation of the total stellar mass. As we will show in a following Figure, the model star formation rates and the balance of stellar populations in the model galaxies are producing realistic colours. On the other hand, as highlighted by the different age estimates of ATLAS 3D galaxies shown in Fig.~(\ref{ages}), the Lick indexes method tends to systematically bias the recovered stellar ages towards younger values, and to increase the scatter. This problem is exacerbated in star-forming galaxies by the decreasing mass-to-light ratio of younger/intermediate stellar populations (see Pforr et al. 2012). 

Fig.~(\ref{ML}) shows the r-band mass-to-light ratio M/L for the model galaxies, split once again in the three evolutionary types and represented here by the $1,3\sigma$ contours, with the same contour colours as Figs. (\ref{atlas}) and (\ref{ages}). The \textit{filled circles} represent the ATLAS 3D galaxies from Cappellari et al. (2013b), colour-coded again with the logarithm of their effective radius (see the colour bar). 

We see an excellent agreement between the model and the ATLAS 3D sample. We notice that model galaxies of decreasing angular momentum move along the mass-M/L relation defined by the data, from high mass and high M/L to low mass and low M/L. We also notice the trend with radius of the ATLAS 3D galaxies, orthogonal to the mass-M/L relation. More compact objects tend to have a higher M/L and a lower mass. 

Fig. (\ref{colormag}) shows the colour-magnitude diagram $(g-i)$ vs $M_i$ for the model galaxies, divided into the same three evolutionary types, and represented here by the $1,3\sigma$ contours, with the same contour colours as Figs. (\ref{atlas}). The circles represent the Gavazzi et al. (2010) SDSS sample, colour-coded in visual morphology: red for ellipticals and S0s, green for disk galaxies (Sbc and later) and blue for disk galaxies with prominent bulges (Sa-Sb).

We notice good agreement between model and data, although the model is on average $\sim 0.1$ dex too red for $M_i > -21$, which is connected with the tendency of the model to overestimate the ages at low masses. Notice that the model galaxies dominated by instability-driven components reproduce the behaviour of the (Sa-Sb) sample, which populates the luminous tip of the blue cloud, fills up the green valley, and reaches up into the red sequence at intermediate luminosities. As discussed in Paper I, secular evolution dominates galaxies at intermediate masses, leading to the formation of instability-driven bulges and affecting the star formation history, reddening their colours and raising their mass-to-light ratio away from those of the disk population.    

\subsection{Stellar populations and angular momentum}

\begin{figure*}
\includegraphics[scale=0.43]{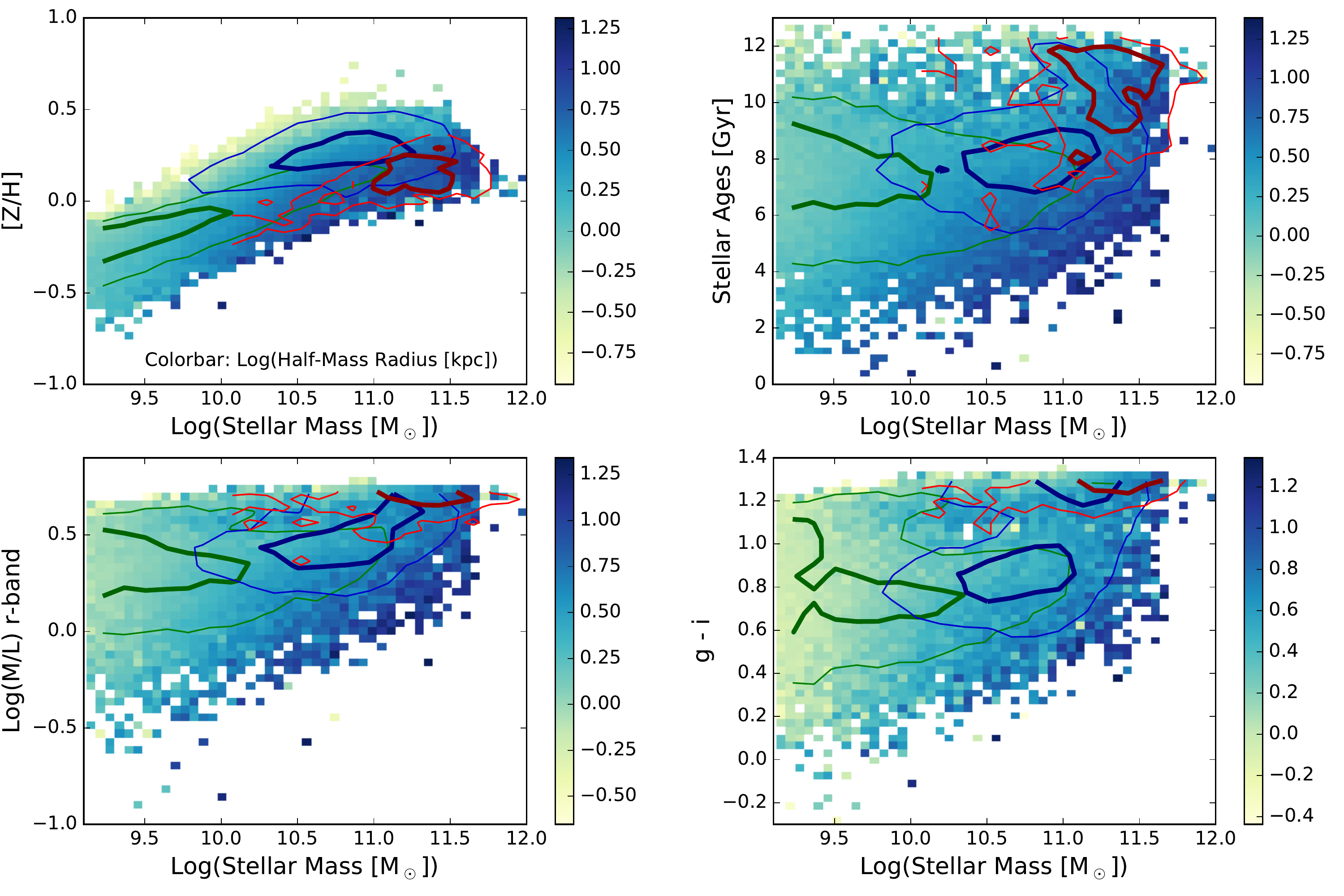}
\caption{Stellar population properties of the model galaxies, as a function of galaxy stellar mass, colour-coded with the galaxy half-mass radius (colorbar). Top left: stellar metallicity. Top right: mass-weighted average age of the stellar populations. Bottom left: mass-to-light ratio in the SDSS-r band (Log solar units). Bottom right: colour $g - i$ (SDSS filters). In all panels, the density contours represent the $1\sigma$ and $2\sigma$ confidence levels for the galaxy evolutionary types described in Figure \ref{atlas}: red for merger-driven, green for disk-dominated, blue for secular-dominated.}
\label{rainbow-rhalf}
\end{figure*}

\begin{figure*}
\includegraphics[scale=0.43]{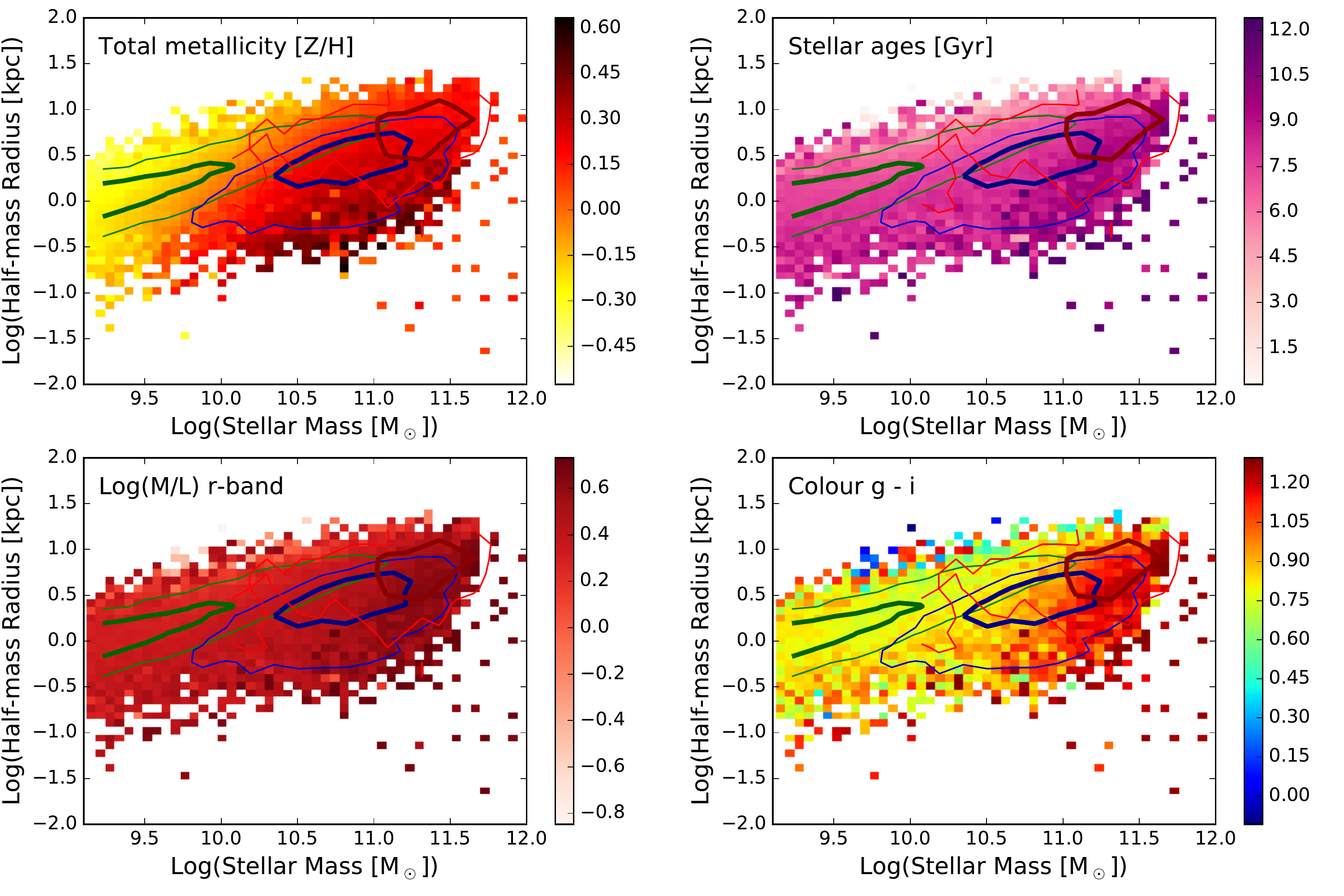}
\caption{Mass-size relation of the model galaxies, colour-coded with stellar population properties (colorbars). Top left panel: total stellar metallicity [Z/H]. Top right: mass-weighted average stellar age (in Gyr). Bottom left: mass-to-light ratio in the SDSS-r band (Log solar units). Bottom right: colour $g - i$ (SDSS filters). In all panels, the density contours represent the $1\sigma$ and $2\sigma$ confidence levels for the galaxy evolutionary types described in Figure \ref{atlas}: red for merger-driven, green for disk-dominated, blue for secular-dominated.}
\label{rainbow-4panels}
\end{figure*}

Now that we have some confidence that the model is producing realistic stellar populations together with a realistic distribution of angular momentum, we proceed to a more detailed analysis of how the latter affects the former. 

Figure (\ref{rainbow-rhalf}) shows a collection of properties of the stellar populations in the model galaxies, as a function of the stellar mass, and colour-coded with the values of the half-mass radius (in Log). The top left panel shows the stellar mass - stellar metallicity relation, the top right panel shows the mass-weighted average age of the stellar populations (in Gyr), the bottom left panel shows the mass-to-light ratio in SDSS r-band, and the bottom-right panel shows the colour $g-i$ in SDSS filters.  In all panels, the contour lines represent the $1\sigma$ and $2\sigma$ confidence levels for the model disks (green), merger-driven objects (dark red) and galaxies with instability-driven components (blue); objects are defined as in Figure \ref{atlas}.

Figure (\ref{rainbow-4panels}) contains the same information as Figure (\ref{rainbow-rhalf}), but this time the mass-size relation is portrayed, colour-coded with the stellar population properties (as indicated in each panel). We include this Figure to facilitate a direct comparison with the results of Cappellari et al. (2013b), where the ATLAS 3D galaxies are portrayed in similar fashion. 
 
All stellar population average properties show a trend with galaxy mass, in the sense that increasing galaxy mass is accompanied by an increase in metallicity and age, an increased mass-to-light ratio and reddening of the colours.
The trends become stronger if only galaxies above $\rm{Log}(M_{\rm{star}}/M_{\odot}) > 10$ are considered. However, apart from the metallicity, the stellar population properties show a very large scatter. This scatter can be deconstructed in terms of both galaxy half-mass radius and galaxy evolutionary type, as described in Fig.(\ref{atlas}). 

At a given galaxy mass, Figs. (\ref{rainbow-rhalf}) and (\ref{rainbow-4panels}) show that more compact objects are more metal-rich, their stellar populations are older, the mass-to-light ratio is higher and the colour is redder. In short, more compact galaxies host more evolved stellar populations. This is also in accord with Atlas 3D results from Cappellari et al. (2013b). This trend is rather sharp in metallicity, and more subtle for the other properties, where it becomes clearer above $\rm{Log}(M_{\rm{star}}/M_{\odot}) > 10$. At lower masses the galaxy population is dominated by disks, and the ongoing star formation in these galaxies diffuses the trend. 

The galaxy evolutionary type sharply defines the location of each galaxy in these diagrams. 

Elliptical galaxies (red) dominate the tip of the mass - size relation (Fig. \ref{rainbow-4panels}). They are mostly composed of the oldest stellar populations, they have the highest mass-to-light ratios, and in the colour - mass diagram they form a distinctive red sequence, and they are not found in the blue cloud (Fig. \ref{rainbow-rhalf}). 

Galaxies with instability-driven bulges, dominated by secular evolution (see Paper I), comprise most of the galaxy population at intermediate masses ($10 > \rm{Log}(M_{\rm{star}}/M_{\odot}) > 11$). Their stellar populations are the most metal-rich, they have intermediate ages and mass-to-light ratios, and in these properties they show a continuity with the disk populations, as expected. Interestingly, above $\rm{Log}(M_{\rm{star}}/M_{\odot}) > 10$ galaxies with instability-driven bulges show a red sequence of their own, and they dominate the green valley as well as the upper blue cloud. In Paper I we discussed how the growth of the instability-driven bulge is connected with the reddening of the galaxy colours towards the green valley and the red sequence. 

Finally, disks dominate the lower mass end of the galaxy population. They show intermediate-to-low ages and mass-to-light ratios, and they constitute most of the blue cloud (although at low masses they show significant scatter in their stellar population properties, reflecting the large variety of merger tree activity of small halos). 

The stellar population properties discussed here are connected to various timescales of the galaxy activity. In particular, the total metallicity and age are probes of the star formation history, while the mass-to-light ratio and the colours are affected by the instantaneous star formation rate. 
The progressive increase of the values of the stellar population properties corresponds to a diminishing galaxy size, and most of all by a shift in galaxy structure from disky to bulgy, through the growth of the instability-driven bulge at intermediate masses, and the dominance of major mergers at high masses.
We can quantify galaxy structure with angular momentum, going from disk galaxies with high angular momentum, to intermediate angular momentum types dominated by instability-driven bulges, to ellipticals with low to null angular momentum. In the model the gas angular momentum determines the star formation rate, and the sequence of mass growth events (including star formation, minor and major mergers) determines the stellar angular momentum. Therefore, there is a direct connection between the galaxy angular momentum and the stellar populations. For a given galaxy mass, lower angular momentum corresponds to more evolved stellar populations. 

\section{Discussion}   

The new semi-analytic model discussed here and in Paper I connects galaxy structure evolution and stellar populations in three ways. 

First, it follows the angular momentum of each gaseous and stellar component (cold gas, stellar disk, instability-driven and merger-driven bulge) self-consistently based on their mass accretion histories. This allows the model to calculate the normalisation and scale length of the exponential density profiles for all components (see Paper I). 

Second, by knowing the gas angular momentum we can self-consistently calculate the star formation rate based on the size and density profile of the gas disk. This connection between the star formation rate and the dynamical state of the gas in the disk, and the corresponding decoupling of the disk structure from the highly-varying spin of the dark matter halo (dominated by interactions with neighbours at its outskirts, as discussed in Paper I) has a stabilising effect on the baryonic cycle. In addition, the evolution of angular momentum and size of the stellar disk is determined by the star formation rate and the structure of the gas disk. 

Third, we model each episode of mass accretion based on the dynamics of the interaction, and the structure of the central galaxy. This implies that a galaxy conserves a memory of its previous dynamical state, and the effects of mass accretion vary depending on the galaxy structure.
This has significant consequences when the mass growth is gradual in time. For instance, in case of minor mergers in real galaxies, new material is accreted peripherally in the galaxy, either in the disk (from where gravitational instabilities transfer mass into the bulge) or at the outskirts of the bulge itself in case of bulge-dominated galaxies. Following this rationale, the new model implementation of minor mergers prevents the stellar content from the satellite, which is typically low in metallicity, to dilute the stellar populations of the bulge. 

The comparison with SAGE (Croton et al. 2016) shows that the new model is more effective at producing self-enrichment. Note that the new star formation law has the effect of creating a bottleneck in the conversion of gas into stars. While in SAGE the gas is uniformly distributed in the disk, in the new model the gas has an exponential density profile, and only the gas inside a critical radius is above the threshold for star formation. But this is the densest portion of gas in the disk, and in both models the star formation rate is proportional to gas density, not mass. To this we add that the density itself is calculated with more realistic disk radii produced in the new model (see paper I). The result is less gas mass converted into stars in a single burst, but with a relatively higher star formation rate, and more bursts of star formation needed to consume a given gas mass. The combined effect is a boost in the production of metals and self-enrichment. 
It is interesting to notice that our angular momentum-based conversion of gas into stars is a complementary approach to models that calculate the star formation rate based on the molecular hydrogen fraction (for instance Blitz \& Rosolowsky 2006, Krumholtz et al. 2012). This fraction depends on gas density or pressure which, like in our model, leads to more intense star formation in denser gas disks. The comparison between these models and ours deserves more investigation, but is beyond the scope of the present work. 

The stellar populations of the model galaxies show a variation with galaxy dynamical structure, in accord with ATLAS 3D data. The dynamical structure
is determined by the evolutionary history: 1) quiescent, mostly in-situ star formation that produces disk galaxies, 2) intense star formation history with an abundance of perturbations in intermediate-richness environments, leading to massive disks and instability-driven bulges (secular-like evolution), and 3) evolution dominated by major mergers (and late-time minor mergers), leading to elliptical galaxies. Each of these galaxy types is characterised by a different mass-size relation, and by an increasing dissipation of angular momentum. In the mass-size relation diagram, they align with the Atlas 3D galaxies classified as disks, fast rotators and slow rotators (Cappellari et al. 2013a). This alignment persists when the stellar populations are investigated through their metallicity, age, colour and mass-to-light ratio (Cappellari et al. 2013b), showing that the model captures the link between assembly history and stellar population properties of the observed sample. 

Model ellipticals in particular, which accumulate most of their mass through major mergers, are in general comprised of old and metal-rich stellar populations. These features highlight the underlying regularity of hierarchical clustering. The most massive of these objects grow on top of large cosmological perturbations, which start to develop early and come fully together at later times. In such over-dense regions evolution is accelerated compared to lower density regions, and star formation rates are high at very high redshifts. These regions are therefore very efficient in producing metals and retaining them in their deep potential wells. 
The overall growth of structure locks away these metal-rich stellar populations in massive ellipticals, so that the densest regions in the universe are the most evolved. Low redshift massive ellipticals are relics of the sites that have produced stars and metals at high rates thoughout cosmic time. 

Note however that this does not imply that massive ellipticals form first, as is the case of the ''red and dead'' scenario. The latter requires that massive old galaxies with high metal content must form at very high redshift, in gigantic bursts of star formation that put almost all the mass in place at very early times, and stop all subsequent activity (see the discussion in Tonini et al. 2012). 

Slow evolution on the other hand, such as that found in galaxies with extended star formation histories and abundance of minor mergers, creates a larger scatter in all stellar population properties. The source of this scatter is to be found in environmental conditions, or the richness of the merger tree, which determines for instance gas infall and the onset of disk instabilities from minor mergers. It can be argued that the speed of the merger tree determines the angular momentum evolution, so the slower the tree, the more angular momentum is retained in the galaxy. 

The connection between stellar populations and dynamical structure shown by the model galaxies supports a scenario of structural quenching. The behaviour of the model galaxies seems to be in accord with recent results (see for instance Ownsworth et al. 2016) that indicate that galaxy structural changes drive the star formation history, and in particular the quenching of star formation. The dissipation of angular momentum in model galaxies, whether in major mergers or during the growth of the instability-driven bulge, seems always to be accompanied by the presence of more evolved stellar populations. In Paper I we noted how the green valley is almost exclusively composed of galaxies with instability-driven bulges. Here we also notice, beside the reddening of the colours, an increase in metallicity, mass-to-light ratio and age.
Angular momentum dissipation could be the cause of an accelerated evolution. During perturbations such as mergers and disk instabilities, gas loses angular momentum and is compressed more efficiently, leading to more violent bursts of star formation than in the case of quiescent evolution. In addition, after such perturbations the galaxy is denser than before (for instance, stars have been transferred from the disk to the bulge through instabilities), creating the conditions for subsequent infalling gas to reach higher densities faster. It is an interesting scenario, and more work is needed to investigate it. In particular, a future paper will be dedicated to a comprehensive statistical analysis of the star formation histories as a function of angular momentum evolution.

This connection emerging between environment, speed of mass accretion, angular momentum evolution and galaxy structure presents us with the possibility to compare the ATLAS 3D classification of fast and slow rotators with the relative incidence of secular versus violent processes in our model galaxies. This is an intriguing possibility, that deserves more work in the future.

\section{Conclusions}

In this work we have studied the properties of the stellar populations of model galaxies produced by the semi-analytic model described in Paper I (Tonini et al. 2016). In this model we follow the evolution of the angular momentum of gas and stars, produce a star formation recipe that depends on the gas disk dynamical structure, treat the effects of mergers depending on the structure of the central galaxy, and evolve the angular momentum of all galaxy components self-consistently, based on the star formation and mass accretion history. The model produces a range of galaxy evolutionary types, and in particular two classes of bulges, namely instability-driven and merger-driven, with separate populations and dynamical properties. 

We have investigated the stellar population properties across the model galaxy population, including metallicity,  colours, age, and mass-to-light ratio in relation to the galaxy structure and evolutionary type. Our results are as follows: 

$\bullet$ The model reproduces the observed stellar mass - stellar metallicity relation in the range $\rm{Log}(M_{\rm{star}}/M_{\odot}) > 10$ for the bulk of the galaxy population. For non-elliptical galaxies the model produces a longer than observed high-metallicity tail. The model also reproduces the relation for Brightest Cluster Galaxies;

$\bullet$ Comparison with the model SAGE shows that the new implementation boosts the efficiency of the baryonic cycle in producing and recycling metals. It also boosts the ability of hierarchical clustering to redistribute them, particularly to merger-driven objects, resulting in agreement with the observed ellipticals metallicity;

$\bullet$ All galaxy components - disks, instability-driven and merger-driven bulges - show a mass-metallicity relation and a high-metallicity tail. Merger-driven bulges show the sharpest stellar mass-metallicity relation, while the other components show increased scatter. The relation for elliptical galaxies stems from the underlying regularity of cosmic structure growth that locks in the metals in the densest regions. For disks and instability-driven bulges, local environment and the richness of the merger tree introduce additional scatter, due to their prologed formation timescales;

$\bullet$ The mass-size relation of galaxies with quiescent assembly histories, with secular-like evolution histories (dominated by intense star formation and minor mergers) and with violent merger-driven histories, align well with the observed galaxies of the Atlas 3D samples of disks, fast rotators and slow rotators respectively; 

$\bullet$ The stellar population properties - metallicity, age, mass-to-light ratio and colour - show that both more massive and more compact galaxies host more evolved stellar populations; 

$\bullet$ The galaxy evolutionary type is the major factor in determining the stellar population properties at a given galaxy mass. A lower angular momentum content, such as that of merger-driven objects, or objects dominated by the growth of an instability-driven bulge, is accompanied by ageing and reddening of the stellar populations, and a boost in metallicity. More evolved stellar populations in the model tend to reside in the objects with the lower angular momentum, in accord with Atlas 3D observations. 
 
The timescales of galaxy growth and the transformations of galaxy dynamical structure are imprinted in the stellar populations. A detailed comparison of the model predictions with observations from large IfU surveys like SAMI (Croom et al. 2012, Allen et al. 2015) and MaNGA offers the scope for future work.

\section*{Acknowledgments}
We wish to thank the anonymous Referee for their useful comments and suggestions. CT wishes to thank Michele Cappellari, Julien Devriendt, Richard McDermid, Eric Bell, Martin Bureau, Nic Scott, Rosa Gonzalez Delgado and Benedetta Vulcani, for the interesting discussions and helpful suggestions. This work was funded by the Australian Research Council through the ARC DP140103498.

\end{document}